\documentclass[12pt]{article}
\usepackage{graphicx}
\usepackage{amsmath}
\usepackage{amssymb}
\usepackage{cite}
\usepackage{geometry}
\usepackage{booktabs}
\usepackage{array}
\usepackage{multirow}
\usepackage{paracol}
\usepackage{longtable}
\usepackage{hyperref}
\usepackage{caption}
\usepackage{float}

\title{\textbf{Autonomous AI for Multi-Pathology Detection in Chest X-Rays: A Multi-Site Study in Indian Healthcare System}}

\author{
    Bargava Subramanian, Dr. Shajeev Jaikumar, Dr. Praveen Shastry \\
    Naveen Kumarasami, Kalyan Sivasailam, Anandakumar D \\
    Keerthana R, Mounigasri M, Kishore Prasath Venkatesh
}

\date{}

\begin{document}

\maketitle

\section*{Abstract}

\paragraph{Study Design} The study outlines the development of an autonomous AI system for chest X-ray (CXR) interpretation, trained on a vast dataset of over 5 million X-rays sourced from healthcare systems across India. This AI system integrates advanced architectures, including Vision Transformers, Faster R-CNN, and various U-Net models (such as Attention U-Net, U-Net++, and Dense U-Net), to enable comprehensive classification, detection, and segmentation of 75 distinct pathologies. To ensure robustness, the study design includes subgroup analyses across age, gender, and equipment type, validating the model's adaptability and performance across diverse patient demographics and imaging environments.

\paragraph{Performance} The AI system achieved up to 98\% precision and over 95\% recall for multi-pathology classification, with stable performance across demographic and equipment subgroups. For normal vs. abnormal classification, it reached 99.8\% precision, 99.6\% recall, and 99.9\% negative predictive value (NPV). Deployed in 17 major healthcare systems in India, including diagnostic centers, large hospitals, and government hospitals. Over the deployment, it processed around 150,000+ scans, averaging 2,000 chest X-rays daily, resulting in reduced reporting times and improved diagnostic accuracy.

\paragraph{Conclusion} The high precision and recall validate the AI’s capability as a reliable tool for autonomous normal/abnormal classification, pathology localization, and segmentation. This scalable AI model addresses diagnostic gaps in underserved areas, optimizing radiology workflows and enhancing patient care across diverse healthcare settings in India.

\paragraph{}

\maketitle

\section*{Introduction}

Chest X-ray (CXR) imaging is an essential diagnostic tool in India, widely used for detecting thoracic diseases, including pneumonia, tuberculosis (TB), lung cancer, and cardiovascular conditions. India bears a significant burden of infectious and respiratory diseases, with TB alone contributing to over a quarter of global cases \cite{Wang et al. 2017}. Millions of CXRs are conducted annually, creating a substantial demand for radiologists, especially given the limitations of traditional interpretation methods. These methods struggle with high workloads, variability in interpretations, and the risk of missed diagnoses, with studies indicating that up to 30\% of abnormalities may go undetected \cite{Prevedello et al. 2017}. This underscores the urgent need for enhanced solutions to improve the accuracy and efficiency of CXR analysis.

The healthcare system in India is further strained by a critical shortage of radiologists. With fewer than 15,000 radiologists serving a population of over 1.4 billion, there is an immense gap between demand and supply \cite{Candemir et al. 2014}. This shortage is particularly pronounced in rural areas, which lack adequate access to diagnostic services, leaving a large portion of the population underserved. The growing burden of diseases such as tuberculosis and post-COVID-19 complications further amplifies the need for timely, scalable diagnostic solutions \cite{Almeida et al. 2017}. Addressing these challenges requires a shift towards AI-driven technologies that can alleviate the workload, standardize reporting quality, and provide quicker patient care.

This paper presents an AI-based approach designed to address these challenges in the Indian healthcare system by enhancing CXR pathology detection capabilities. The system is capable of detecting 75 distinct pathologies, from common infections to complex thoracic conditions, using a combination of Vision Transformers for classification, Faster R-CNN for detection, and UNet for segmentation \cite{Esteva et al. 2019}. This multi-layered approach enables precise identification of abnormalities while automating the detection and reporting process. We begin by outlining the challenges in CXR reporting within India and describe how AI is uniquely positioned to tackle these issues \cite{Lundervold et al. 2019}. The paper also elaborates on the methodology, covering the system architecture and workflow from input to pathology detection, aimed at providing accurate and actionable insights \cite{Jalloul et al. 2024}.

The results section highlights the effectiveness of this AI solution, showing precision rates up to 97\% and recall exceeding 95\%, demonstrating its ability to enhance diagnostic workflows significantly. By reducing reporting times by up to 50\%, the AI-driven approach can address the high demand for CXR interpretation in India, thereby improving patient care and clinical outcomes \cite{Olawade et al. 2023}. This integration of AI into clinical workflows offers an opportunity to bridge the diagnostic gap, particularly in underserved areas, supporting healthcare professionals, enhancing diagnostic accuracy, and ultimately contributing to better patient outcomes across India \cite{Tang et al. 2020}.

\maketitle

\section*{Methodology}

\subsection*{AI System Overview} 
The AI system developed for this study is a computer-aided detection (CAD) tool designed for the identification and differentiation of various radiological abnormalities present in chest X-rays (CXRs). This system incorporates multiple deep-learning algorithms, each tailored to detect specific pathologies, covering a comprehensive range of thoracic conditions(Ronneberger et al., 2015). The models were trained on a large-scale dataset consisting of over 5 million CXR images, with expert radiologist annotations used for supervised learning(Firdiantika \& Jusman, 2022). The AI system aims to detect abnormalities such as lung nodules, pleural effusion, pneumothorax, cardiomegaly, consolidation, fibrosis, hilar enlargement, rib fractures, etc.

\begin{center}
\begin{table}[H]
\begin{tabular}{|p{0.3\textwidth}|p{0.3\textwidth}|p{0.3\textwidth}|}
\hline
Alveolar Lung Opacity & Foreign Body - Intercostal & Nodule \\
\hline
Atelectasis & Foreign Body - Nasogastric Tube & Old Healed Clavicle Fracture \\
\hline
Azygous Lobe & Foreign Body - Nasojejunal Tube & Old Rib Fracture \\
\hline
Bifid Rib & Foreign Body - Pacemaker & Old Tuberculosis \\
\hline
Bronchiectasis & Foreign Body - Pigtail Catheter & Pericardial Cyst \\
\hline
Bullous Emphysema & Foreign Body - Spinal Fusion & Pleural Calcification \\
\hline
Cardiomegaly & Foreign Body - Sternal Sutures & Pleural Effusion \\
\hline
Cavity & Foreign Body - Tracheostomy Tube & Pleural Plaque \\
\hline
Cervical Rib & Hilar Lymphadenopathy & Pleural Thickening \\
\hline
Clavicle Fracture & Hilar Prominence & Pneumonia \\
\hline
Clavicle Fracture with PO & Humerus Fracture & Pneumoperitoneum \\
\hline
Consolidation & Humerus Post OP & Pneumothorax \\
\hline
Dextrocardia & Hydro Pneumothorax & Prominent Bronchovascular Markings \\
\hline
Dextrocardia with situs inversus & Hypoplastic Rib & Pulmonary Edema \\
\hline
Diaphragmatic Hump & Interstitial Lung Disease & Reticulo-nodular Appearance \\
\hline
Elevated Diaphragm & Interstitial Lung Opacity & Rib Fracture \\
\hline
Esophageal Stent & Lobe Collapse & Scapula Fracture \\
\hline
Fibrosis & Lung Collapse & Scoliosis \\
\hline
Fissural Thickening & Lung Mass & Subcutaneous Emphysema \\
\hline
Flattened Diaphragm & Lymph Node Calcification & Surgical Staples \\
\hline
Foreign Body - Cardiac Valves & Mastectomy & Thyroid Lesion \\
\hline
Foreign Body - Chemoport & Mediastinal Mass & Tracheal and Mediastinal Shift \\
\hline
Foreign Body - Chest Leads & Mediastinal Shift & Tracheal Shift \\
\hline
Foreign Body - CV Line & Mediastinal Widening & Tuberculosis \\
\hline
Foreign Body - Endotracheal tube & Milliary Tuberculosis & Unfolding of Aorta \\
\hline

\end{tabular}
\caption{List of 75 Pathologies}
\end{table}
\end{center}

The detection system employs a stepwise approach to analyze CXRs, focusing on both the identification of specific abnormalities and the overall classification of findings. Initially, the system processes the input image to classify it as whether it‘s a valid CXR or not. Following this initial classification, the model performs detailed detection to identify and localize specific pathologies within the CXR, such as lung nodules, pleural effusion, pneumothorax, and other abnormalities \cite{Ahmed et al. 2024}.

In this study, the primary focus was on the comprehensive detection of various pathologies present in each CXR, rather than just distinguishing between normal and abnormal \cite{Nam et al. 2019}. By doing so, the system enables a more thorough analysis of each radiograph, providing granular details about identified pathologies. This focus on detailed detection enhances the accuracy of the diagnostic process and aids radiologists in making informed decisions, ultimately contributing to better patient outcomes \cite{Albertini et al. 2017}.

\subsection*{Dataset}

This study utilized a dataset of 5,003,742 chest X-ray (CXR) scans gathered from various healthcare facilities. The dataset was split into three subsets: Training, Live Clinical Trial, and Live Clinical Deployment, with slightly varied proportions to support model development, validation, and real-world application.

\begin{center}
    \parbox{0.6\textwidth}{%
        Training Set: 3,997,891 scans \\ 
        Live Clinical Trial: 855,482 scans \\ 
        Live Clinical Deployment: 157,369 scans
    }
\end{center}

\vspace{2cm}
\subsection*{Age Group Distribution}
Scans were distributed across age groups to capture demographic diversity:
\begin{table}[h!]
\centering
\begin{tabular}{|p{0.15\textwidth}|p{0.15\textwidth}|p{0.15\textwidth}|p{0.15\textwidth}|p{0.15\textwidth}|}
\hline
\textbf{Age Group} & \textbf{Total Scans} & \textbf{Training Set} & \textbf{Live Clinical Trial} & \textbf{Live Clinical Deployment} \\
\hline
Under 18 & 691,487 & 600,208 & 75,732 & 15,547 \\
\hline
18-40 & 1,779,807 & 1,404,071 & 327,429 & 48,307 \\
\hline
40-60 & 1,424,936 & 1,201,381 & 189,765 & 33,790 \\
\hline
60-75 & 1,187,200 & 1,004,752 & 151,034 & 31,414 \\
\hline
75+ & 933,312 & 793,479 & 111,522 & 28,311 \\
\hline
\end{tabular}
\caption{Scans distribution based on Age Group}
\end{table}

\newpage
\subsection*{Gender Distribution}
The dataset maintained a balanced gender distribution:
\begin{table}[h!]
\centering
\begin{tabular}{|p{0.15\textwidth}|p{0.15\textwidth}|p{0.15\textwidth}|p{0.15\textwidth}|p{0.15\textwidth}|}
\hline
\textbf{Gender} & \textbf{Total Scans} & \textbf{Training Set} & \textbf{Live Clinical Trial} & \textbf{Live Clinical Deployment} \\
\hline
Male & 3,338,809 & 2,799,453 & 453,688 & 85,668 \\
\hline
Female & 2,677,933 & 2,204,438 & 401,794 & 71,701 \\
\hline
\end{tabular}
\caption{Scans distribution based on Gender}
\end{table}

\subsection*{Manufacturer Type Distribution}
Scans were categorized by equipment manufacturer to account for variability in imaging conditions:
\begin{table}[h!]
\centering
\begin{tabular}{|p{0.18\textwidth}|p{0.15\textwidth}|p{0.15\textwidth}|p{0.15\textwidth}|p{0.15\textwidth}|}
\hline
\textbf{Manufacturer} & \textbf{Total Scans} & \textbf{Training Set} & \textbf{Live Clinical Trial} & \textbf{Live Clinical Deployment} \\
\hline
GE Healthcare & 1,866,094 & 1,549,878 & 263,487 & 52,729 \\
Siemens & 1,423,256 & 1,152,414 & 230,532 & 40,310 \\
Philips & 1,395,340 & 1,158,413 & 203,287 & 33,640 \\
Other Manufacturers & 1,332,052 & 1,143,186 & 158,176 & 30,690 \\
\hline
\end{tabular}
\caption{Scans distribution based on Manufacturer Type}
\end{table}

\subsection*{Equipment Type Distribution}
Scans were categorized by equipment type to account for variability in imaging conditions:
\begin{table}[h!]
\centering
\begin{tabular}{|p{0.15\textwidth}|p{0.15\textwidth}|p{0.15\textwidth}|p{0.15\textwidth}|p{0.15\textwidth}|}
\hline
\textbf{Machine Type} & \textbf{Total Scans} & \textbf{Training Set} & \textbf{Live Clinical Trial} & \textbf{Live Clinical Deployment} \\
\hline
CR & 4,153,207 & 3,502,723 & 556,063 & 94,421 \\
DR & 1,863,535 & 1,501,168 & 299,419 & 62,948 \\
\hline
\end{tabular}
\caption{Scans by Machine Type}
\end{table}

This dataset composition, with varied proportions across training, trial, and deployment, ensures comprehensive model training, rigorous validation in clinical trials, and real-world testing, supporting reliable application across diverse demographics and equipment settings in clinical workflows. To maintain compliance with ethical guidelines and data protection standards, all patient data underwent rigorous anonymization, ensuring complete privacy by removing identifiable patient information before use in this study, in compliance with HIPAA.

\subsection*{Distinct Quality Challenges in Indian CXR Datasets}
In the context of India, there are unique challenges that significantly impact the quality and diagnostic accuracy of CXR imaging. These include rotations, artifacts, suboptimal images, etc.. Rotational issues in images may result from improper patient positioning, while artifacts can be introduced due to equipment limitations or external objects during imaging. Suboptimal images, often a result of inadequate imaging conditions or limited resources in some facilities, can lead to difficulties in detecting pathologies accurately.
To address these challenges, the dataset included a diverse range of CXR images, spanning from below-average quality to high-quality images. This range allowed the AI model to be trained on various levels of image quality, making it more robust and capable of handling live clinical diagnosis scenarios where image quality may vary significantly. By incorporating these lower-quality images into training, the model becomes better equipped to identify abnormalities effectively across different imaging conditions, enhancing its utility in both urban and rural healthcare settings in India.

\maketitle

\section*{Architecture}

The architecture of the system is divided into multiple phases, including the Annotation Phase, Analysis and Detection Phase. Each of these phases plays a critical role in ensuring accurate and efficient detection of pathologies in chest X-rays (CXRs). Below is a detailed explanation of each phase:

\subsection*{Annotation Phase}

\subsubsection*{Dataset Segregation}

The annotation process forms the foundational step of the overall workflow. It begins by segregating chest radiograph images into unlabelled and labeled datasets \cite{Allaouzi & Ben Ahmed 2019}. This data is further categorized into Posterior-Anterior (PA) views and Anterior-Posterior (AP) views to provide distinct training inputs, as these views are critical for detecting pathologies accurately \cite{Wang et al. 2020}. The annotation phase involves handling both labeled and unlabelled data to ensure a comprehensive and diverse dataset that can enhance model generalizability \cite{Litjens et al. 2017}.

\subsubsection*{Dataset Selection}
After the dataset segregation, the images proceed to dataset selection, where the appropriate subset of data is chosen based on the requirements of the training task \cite{Sheng et al. 2015}. This careful selection helps ensure that only relevant images are processed, which subsequently improves the performance of downstream tasks \cite{Tahmoresnezhad & Hashemi 2016}.

\subsubsection*{Dataset Pre-processing}
Once the dataset is selected, the next step is pre-processing. Pre-processing involves adjusting image quality, size, and other properties to maintain uniformity across the dataset \cite{Frid-Adar et al. 2018}. This consistency in image quality is essential for the model to extract relevant features efficiently and ensures that the CXR images are well-prepared for further analysis \cite{Wu et al. 2023}. This process also plays a role in minimizing noise and enhancing the clarity of key anatomical features \cite{Castiglioni et al. 2021}.

\subsubsection*{Cross Teaching}
To further leverage the unlabelled data, the cross-teaching phase is applied. Cross-teaching is a semi-supervised learning approach where multiple models iteratively train each other using both labeled and unlabelled data, enhancing generalization and robustness\cite{Schlemper et al. 2019}. By integrating labeled and unlabelled data during the learning phase, the system is capable of utilizing a more extensive dataset, ultimately improving detection performance \cite{Lee et al. 2019}.

\subsection*{Training Phase}
The training phase begins with preprocessing and initial classification steps that prepare each image for detailed pathology detection. The input chest X-ray images, originally in DICOM format, are converted to JPEG using \textit{pydicom} for compatibility with deep-learning models. The architecture is designed to handle high variability in imaging quality by incorporating a multi-resolution analysis approach, enabling the model to capture a range of details essential for accurate classification and detection.

\subsubsection*{Initial Classification and Preliminary Verification with Vision Transformers}
Each image enters the training pipeline through a Vision Transformer (ViT) model, where initial classifications and sanity checks are performed. These checks include:

\begin{itemize}
    \item \textbf{Image Verification}: The model verifies if the image is an X-ray, filtering out non-X-ray images..
    
    \item \textbf{Chest X-Ray Identification}: The next classification confirms if the X-ray is specifically a chest X-ray, distinguishing it from other anatomical regions (e.g., extremities, abdomen).
    
    \item \textbf{View Classification}: 
    The model first analyzes the chest X-ray to classify it as either Posterior-Anterior (PA) or Anterior-Posterior (AP), ensuring accurate interpretation. This classification is essential as PA and AP views differ in anatomical perspective, impacting pathology detection.
    
    \item \textbf{Rotation Correction with Keypoint Detection}: The model detects specific anatomical landmarks, such as the clavicles and spinous process, to correct any rotational misalignment in the X-ray. By aligning these key points—especially focusing on the relative positions of the clavicles and spinous process—the model can accurately compute and apply the necessary rotation adjustments, ensuring that all images have a consistent, upright orientation.
\end{itemize}

\subsubsection*{Normal and Abnormal Classification with Multi-Resolution Analysis}
Following the sanity checks, each radiograph is classified as normal or abnormal using a Vision Transformer. In this phase, multi-resolution analysis is a critical component of training, allowing the model to analyze images at different pixel resolutions to capture features across varying levels of detail:

\begin{itemize}
    \item \textbf{224x224 pixels}: This lower resolution provides a broader view of the structure, suitable for identifying larger abnormalities and general patterns within the chest X-ray.
    
    \item \textbf{320x320 pixels}: At this intermediate resolution, the model gains access to finer details, making it more sensitive to subtle features that may not be visible at lower resolutions.
    
    \item \textbf{512x512 pixels}: This higher resolution is used to capture intricate details and small abnormalities, essential for detecting minor pathologies or subtle changes within the image.
\end{itemize}

\subsubsection*{Output Layer Structuring}
At the final layer of the Vision Transformer used for normal vs. abnormal classification, a 2-class softmax output layer is used. This layer focuses on the binary labels: \textbf{Normal} and \textbf{Abnormal}. This simplified output reduces complexity at the classification step, ensuring that the model remains focused on the essential diagnostic categories.

Each resolution is trained separately on the dataset, and the model predictions from each resolution are then combined using an ensemble averaging technique. This ensemble approach integrates predictions from all three resolutions, enhancing the robustness and accuracy of the normal/abnormal classification. This ensemble approach leverages the unique spatial and structural information captured at each resolution, effectively balancing the detection of both large-scale anomalies and fine-grained pathological features.

This multi-resolution training strategy significantly enhances the Vision Transformer’s ability to distinguish between normal and abnormal cases, supporting more accurate downstream pathology detection and segmentation.

\subsection*{Analysis and Detection Phase}
In the analysis and detection phase, abnormal images are processed through a specialized model stack optimized for comprehensive pathology identification and segmentation. This stack includes a Faster R-CNN for detecting pathology locations and a family of U-Net architectures (Attention U-Net, U-Net++, Dense U-Net) for high-precision segmentation.

\subsubsection*{Pathology Detection with Faster R-CNN}
The Faster R-CNN model detects and localizes abnormalities by generating bounding boxes around potential pathology regions. Key parameter configurations include:

\begin{itemize}
    \item \textbf{Anchor Boxes}: Configured with sizes of 128, 256, and 512 pixels and aspect ratios of 1:1, 2:1, and 1:2 to cover a wide range of pathology sizes.
    
    \item \textbf{Region Proposal Network (RPN)}:
    \begin{itemize}
        \item \textbf{NMS Threshold}: 0.7 to eliminate highly overlapping proposals.
        \item \textbf{Top Proposals}: 2,000 per image during training and 300 during inference for optimal computational efficiency.
    \end{itemize}
    
    \item \textbf{Bounding Box Regression}:
    \begin{itemize}
        \item \textbf{Smooth L1 Loss} with $\beta = 1.0$ to handle localization adjustments without being overly sensitive to small shifts.
        \item \textbf{Bounding Box $\Delta$ Parameters}: Set to $\Delta_x = \Delta_y = 0.1$ and $\Delta_w = \Delta_h = 0.2$ to adjust bounding box predictions for chest X-ray anatomy.
    \end{itemize}
\end{itemize}

\subsubsection*{Segmentation with the U-Net Family (Attention U-Net, U-Net++, Dense U-Net)}
After bounding box localization, each abnormal region is further segmented using a combination of U-Net variants, specifically Attention U-Net, U-Net++, and Dense U-Net. These architectures are chosen for their unique capabilities to improve segmentation accuracy in complex medical images.

\paragraph{1. Attention U-Net}
The Attention U-Net model enhances the base U-Net with attention gates that focus on relevant regions, suppressing irrelevant background features. Key configurations include:

\begin{itemize}
    \item \textbf{Attention Gate Parameters}:
    \begin{itemize}
        \item \textbf{Gate Activation}: Sigmoid with a threshold of 0.5 to selectively highlight pathology-relevant regions.
        \item \textbf{Inter-Channel Weighting}: Emphasizes pathological features by learning spatially adaptive weights, useful in suppressing noise from surrounding anatomical structures.
    \end{itemize}
    
    \item \textbf{Encoder-Decoder Structure}:
    Depth of five levels, with initial filter size of 64, doubling at each level (64, 128, 256, 512, 1024).
\end{itemize}

\paragraph{2. U-Net++}
U-Net++ improves segmentation precision by utilizing densely connected skip pathways and redesigned skip connections, allowing more detailed information transfer between the encoder and decoder.

\begin{itemize}
    \item \textbf{Nested Architecture}: Employs dense skip connections across intermediate layers, allowing the network to learn fine-grained details that enhance segmentation accuracy for complex pathologies.
    \item \textbf{Filter Sizes and Depth}: Five levels, with a base filter size of 64, increasing at each level.
\end{itemize}

\paragraph{3. Dense U-Net}
The Dense U-Net model leverages densely connected layers within the encoder and decoder, enabling feature reuse and better gradient flow, which is particularly useful for handling images with high variability in pathological regions.

\begin{itemize}
    \item \textbf{Dense Connections}: Each layer within the encoder and decoder receives input from all preceding layers, preserving features across all stages and improving the model’s ability to capture details in complex structures.
    \item \textbf{Filter Configuration}: Initial filter size of 32, with four dense blocks, each with growth rate $k = 12$ to control the number of features per layer.
\end{itemize}

\subsubsection*{Training and Inference Parameters for the U-Net Family}
\begin{itemize}
    \item \textbf{Batch Size}: Set to 8 for efficient memory usage while maintaining stable gradients.
    \item \textbf{Learning Rate}: 0.0005 with decay factor 0.9, adjusting every 15 epochs.
    \item \textbf{Dropout Rate}: 0.3 applied in each decoder block to prevent overfitting, given the variability in chest X-ray images.
    \item \textbf{Post-Processing Threshold}: Softmax threshold set at 0.5 across all U-Net outputs to retain high-probability regions as pathologies.
\end{itemize}

\subsection*{End-to-End Workflow}

\begin{itemize}
    \item \textbf{Input Validation and Preliminary Verification}: JPEG-formatted DICOM images undergo validation for modality, chest X-ray type, view identification, and rotation correction.
    
    \item \textbf{Normal/Abnormal Classification with Multi-Resolution Analysis}: A Vision Transformer classifies images at multiple resolutions (224x224, 320x320, and 512x512), followed by ensemble averaging for final output.
    
    \item \textbf{Pathology Detection}: Faster R-CNN generates bounding boxes for each detected pathology.
    
    \item \textbf{Segmentation by U-Net Family}: The Attention U-Net, U-Net++, and Dense U-Net models process each region to produce pixel-level segmentations, enhancing interpretability with high spatial accuracy.
\end{itemize}

\begin{figure}[H] 
    \centering
    \includegraphics[width=1.1\textwidth]{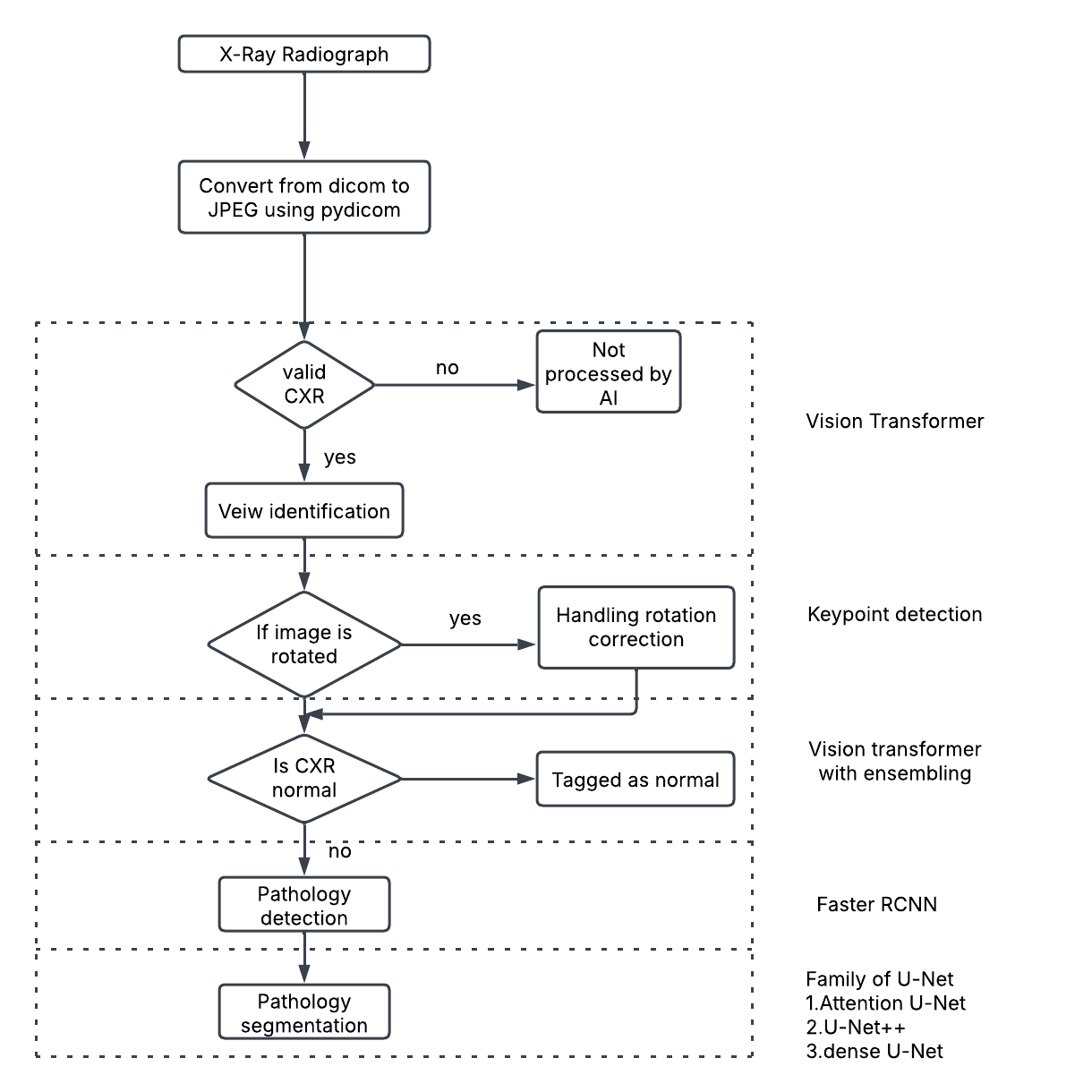} 
    \caption{Workflow Architecture}
    \label{fig:workflow_architecture}
\end{figure}

\maketitle

\section*{Evaluation Metrics: }

The performance of the chest X-ray analysis system was evaluated comprehensively using a set of metrics for both classification and detection tasks to understand its effectiveness in real clinical environments.
For the classification component, we used Negative Predictive Value (NPV) and Positive Predictive Value (PPV) to gauge the model's accuracy in classifying chest X-rays as normal or abnormal. The Positive Percent Agreement (PPA) and Negative Percent Agreement (NPA) metrics were also calculated to determine how well the model's predictions aligned with radiologists' evaluations. These metrics, along with the associated 95\% confidence intervals, provide a robust measure of the system’s accuracy and reliability in distinguishing between normal and abnormal cases.

Moving beyond classification, the detection capability of the model was evaluated using Precision, Recall, and Intersection over Union(IoU) metrics. Precision indicates how many of the abnormalities identified by the model were true positives, while recall measures the ability of the model to identify all existing abnormalities in the chest X-rays. IoU was utilized to evaluate the overlap between predicted regions of interest and the actual ground truth, offering a quantitative assessment of the model's localization accuracy.
Performance metrics for all 75 detected pathologies were documented, providing an in-depth view of how well the system performed for each specific condition. The metrics for precision, recall for each of these pathologies are presented in the table below, showcasing the AI's proficiency in both identifying and accurately localizing abnormalities. These results highlight the system's ability to support clinical workflows, enhancing both the accuracy and speed of radiology reporting.

In general, the combination of classification and detection metrics provides a complete picture of the performance of the model, illustrating its potential to serve as a reliable support tool for radiologists in everyday clinical practice.

\begin{longtable}{|p{0.5\textwidth}|p{0.15\textwidth}|p{0.15\textwidth}|p{0.15\textwidth}|}
\hline
\textbf{Pathology} & \textbf{AUC} & \textbf{Precision (\%)} & \textbf{Recall (\%)} \\
\hline
\endfirsthead

\hline
\textbf{Pathology} & \textbf{AUC} & \textbf{Precision (\%)} & \textbf{Recall (\%)} \\
\hline
\endhead

\hline \multicolumn{4}{r}{\textit{Continued on next page}} \\ \hline
\endfoot

\hline
\caption{Performance Metrics for Detected Pathologies} \label{table:performance_metrics} \\
\endlastfoot
Alveolar Lung Opacity & 0.97 & 97.40 & 95.80 \\
\hline
Atelectasis & 0.98 & 99.40 & 97.40 \\
\hline
Azygous Lobe & 0.99 & 99.31 & 99.29 \\
\hline
Bifid Rib & 0.99 & 95.79 & 94.20 \\
\hline
Bronchiectasis & 0.97 & 98.60 & 98.50 \\
\hline
Bullous Emphysema & 0.97 & 98.09 & 95.22 \\
\hline
Cardiomegaly & 0.96 & 96.50 & 95.90 \\
\hline
Cavity & 0.98 & 98.09 & 97.80 \\
\hline
Cervical Rib & 0.95 & 95.60 & 95.00 \\
\hline
Clavicle Fracture & 0.97 & 96.90 & 95.22 \\
\hline
Clavicle Fracture with PO & 0.98 & 96.90 & 99.40 \\
\hline
Consolidation & 0.96 & 98.09 & 94.20 \\
\hline
Dextrocardia & 0.98 & 97.00 & 98.30 \\
\hline
Dextrocardia with situs inversus & 0.99 & 98.40 & 93.00 \\
\hline
Diaphragmatic Hump & 0.97 & 97.40 & 96.80 \\
\hline
Elevated Diaphragm & 0.98 & 95.70 & 100.00 \\
\hline
Esophageal Stent & 0.96 & 98.00 & 99.30 \\
\hline
Fibrosis & 0.99 & 98.00 & 99.30 \\
\hline
Fissural Thickening & 0.98 & 98.48 & 98.00 \\
\hline
Flattened Diaphragm & 0.99 & 99.31 & 100.00 \\
\hline
Foreign Body - Cardiac Valves & 0.99 & 99.31 & 99.29 \\
\hline
Foreign Body - Chemoport & 0.97 & 97.90 & 95.40 \\
\hline
Foreign Body - Chest Leads & 0.98 & 98.60 & 97.80 \\
\hline
Foreign Body - CV Line & 0.97 & 98.09 & 95.22 \\
\hline
Foreign Body - ETT & 0.98 & 98.73 & 96.59 \\
\hline
Foreign Body - ICD & 0.95 & 95.70 & 95.10 \\
\hline
Foreign Body - Nasojejunal Tube & 0.95 & 99.31 & 99.29 \\
\hline
Foreign Body - NG Tube & 0.96 & 96.30 & 95.90 \\
\hline
Foreign Body - Pacemaker & 0.98 & 98.48 & 98.50 \\
\hline
Foreign Body - Pigtail Catheter & 0.97 & 98.40 & 95.70 \\
\hline
Foreign Body - Spinal Fusion & 1.00 & 100.00 & 99.70 \\
\hline
Foreign Body - Sternal Sutures & 1.00 & 100.00 & 99.80 \\
\hline
Foreign Body - Tracheostomy Tube & 0.96 & 95.70 & 97.00 \\
\hline
Hilar Lymphadenopathy & 0.99 & 98.50 & 95.00 \\
\hline
Hilar Prominence & 0.98 & 99.31 & 99.29 \\
\hline
Humerus Fracture & 0.99 & 99.31 & 98.00 \\
\hline
Humerus Post OP & 0.97 & 98.25 & 96.50 \\
\hline
Hydro Pneumothorax & 0.99 & 98.40 & 97.40 \\
\hline
Hypoplastic Rib & 0.99 & 97.80 & 94.70 \\
\hline
Interstitial Lung Disease & 0.98 & 99.31 & 99.29 \\
\hline
Interstitial Lung Opacity & 0.98 & 98.60 & 97.80 \\
\hline
Lobe Collapse & 0.96 & 96.00 & 95.80 \\
\hline
Lung Collapse & 0.97 & 97.60 & 95.90 \\
\hline
Lung Mass & 0.97 & 97.40 & 96.50 \\
\hline
Lymph Node Calcification & 0.98 & 97.80 & 96.00 \\
\hline
Mastectomy & 0.97 & 98.60 & 97.80 \\
\hline
Mediastinal Mass & 0.96 & 95.90 & 95.00 \\
\hline
Mediastinal Shift & 0.97 & 98.10 & 96.90 \\
\hline
Mediastinal Widening & 0.98 & 98.09 & 95.22 \\
\hline
Milliary Tuberculosis & 0.98 & 98.70 & 95.22 \\
\hline
Nodule & 1.00 & 98.48 & 100.00 \\
\hline
Old Healed Clavicle Fracture & 0.98 & 98.30 & 97.50 \\
\hline
Old Rib Fracture & 0.98 & 100.00 & 100.00 \\
\hline
Old TB & 0.99 & 98.48 & 97.00 \\
\hline
Pericardial Cyst & 0.95 & 96.70 & 95.80 \\
\hline
Pleural Calcification & 0.99 & 99.30 & 97.30 \\
\hline
Pleural Effusion & 0.97 & 97.30 & 96.50 \\
\hline
Pleural Plaque & 0.99 & 99.20 & 98.50 \\
\hline
Pleural Thickening & 0.96 & 96.80 & 95.40 \\
\hline
Pneumonia & 0.97 & 98.09 & 97.80 \\
\hline
Pneumoperitoneum & 0.98 & 98.73 & 96.59 \\
\hline
Pneumothorax & 0.95 & 95.79 & 95.20 \\
\hline
Prominent Bronchovascular Markings & 0.98 & 98.60 & 97.80 \\
\hline
Pulmonary Edema & 0.95 & 96.50 & 94.00 \\
\hline
Reticulo-nodular Appearance & 0.97 & 99.20 & 97.60 \\
\hline
Rib Fracture & 0.98 & 98.20 & 97.60 \\
\hline
Scapula Fracture & 0.96 & 98.50 & 96.30 \\
\hline
Scoliosis & 0.98 & 97.30 & 96.40 \\
\hline
Subcutaneous Emphysema & 0.99 & 98.00 & 99.30 \\
\hline
Surgical Staples & 0.99 & 95.79 & 94.20 \\
\hline
Thyroid Lesion & 0.97 & 96.90 & 99.40 \\
\hline
Tracheal and Mediastinal Shift & 0.97 & 95.79 & 96.00 \\
\hline
Tracheal Shift & 0.98 & 96.90 & 99.40 \\
\hline
Tuberculosis & 0.97 & 97.60 & 100.00 \\
\hline
Unfolding of aorta & 0.98 & 98.60 & 99.29 \\
\hline
\end{longtable}

\begin{figure}[H] 
    \centering
    \includegraphics[width=1\textwidth]{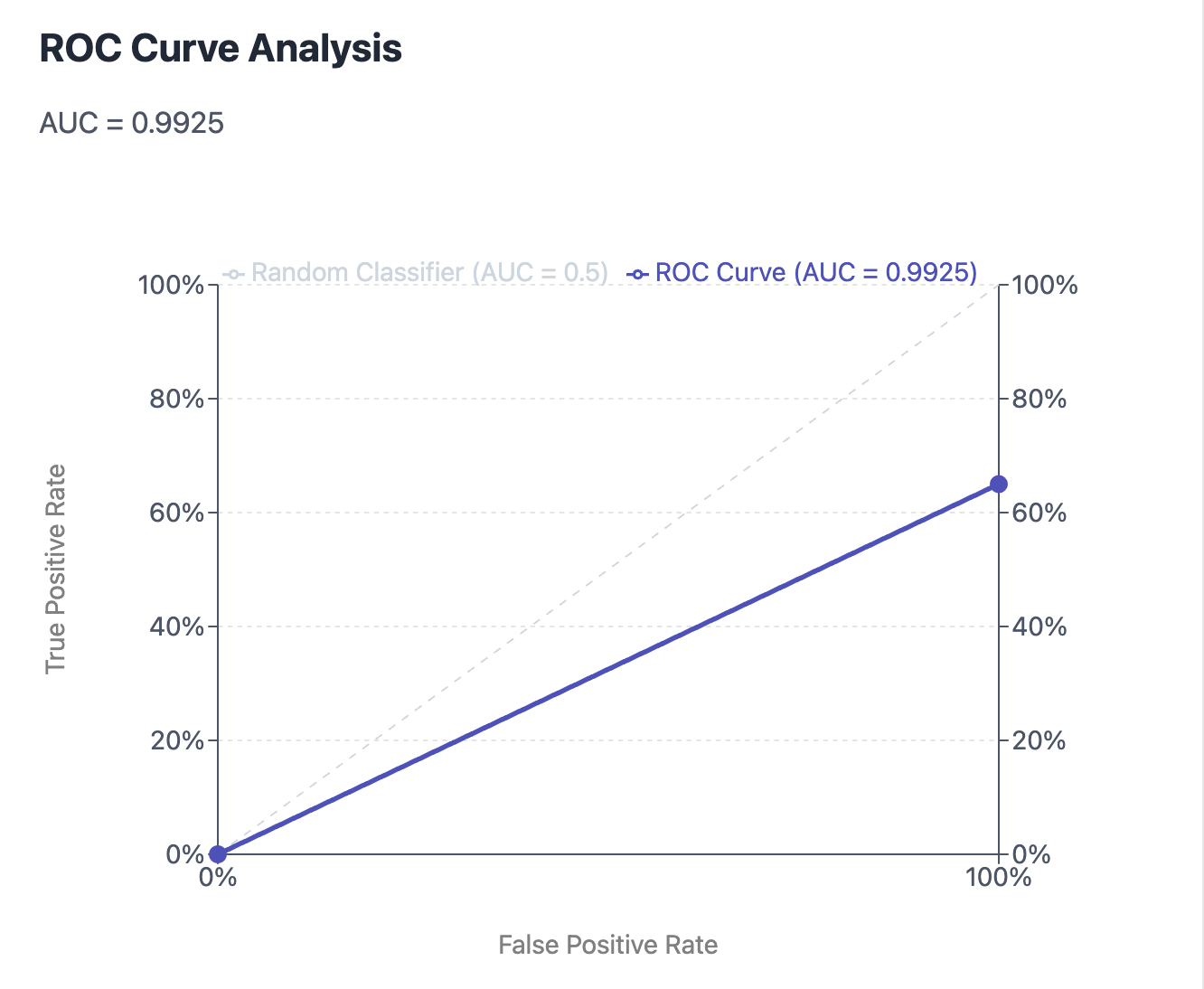} 
    \caption{AUC Curve for Normal/Abnormal Classifier}
    \label{fig:auccurve}
\end{figure}

\begin{figure}[H] 
    \centering
    \includegraphics[width=1.1\textwidth]{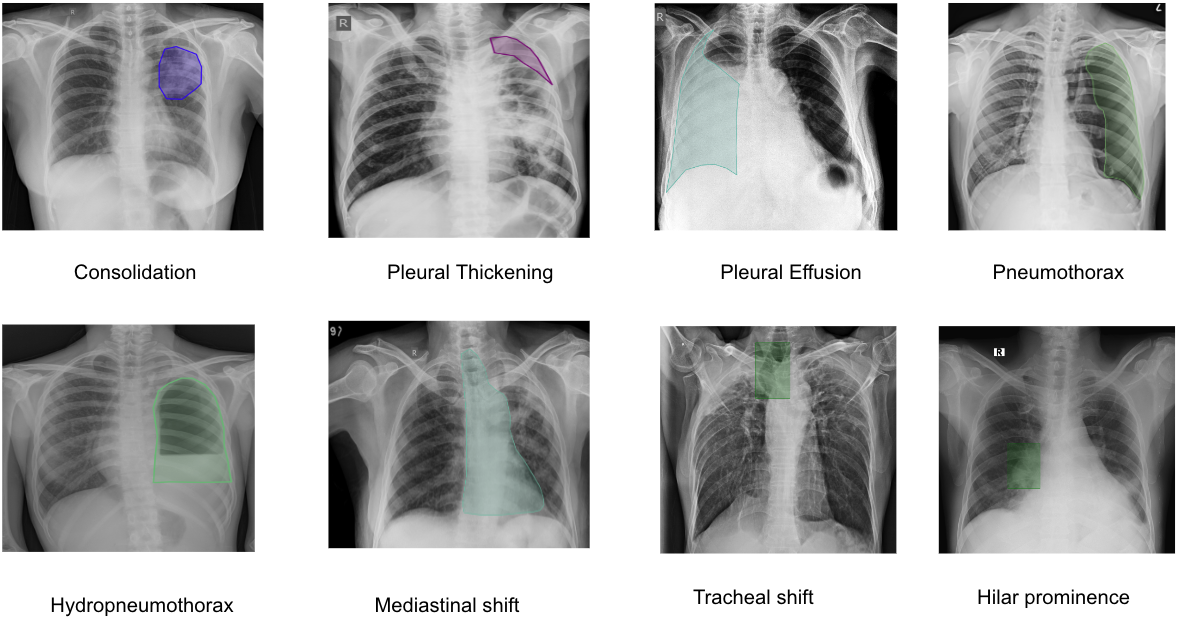} 
    \label{fig:detection1}
\end{figure}

\begin{figure}[H] 
    \centering
    \includegraphics[width=1.1\textwidth]{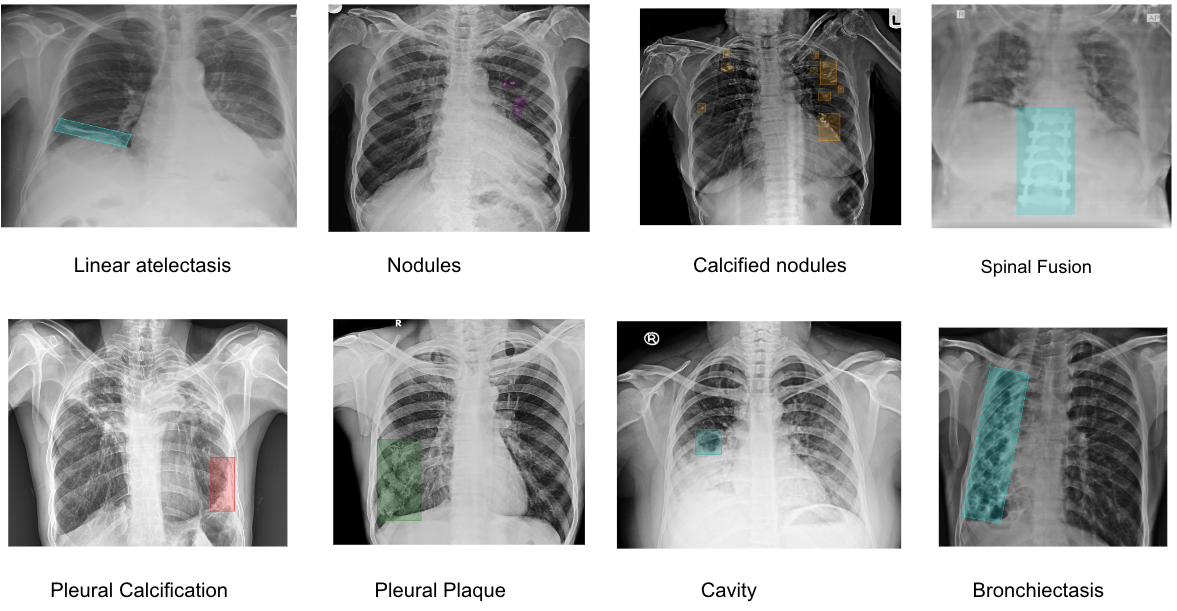} 
    \label{fig:detection2}
\end{figure}

\begin{figure}[H] 
    \centering
    \includegraphics[width=1.1\textwidth]{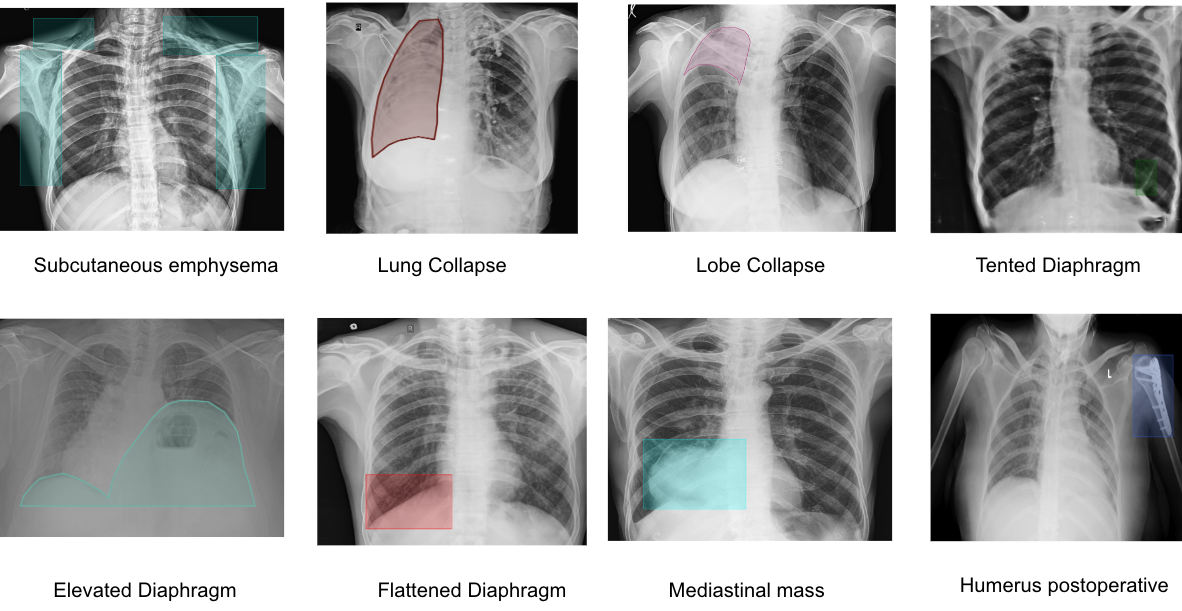} 
    \label{fig:detection3}
\end{figure}

\begin{figure}[H] 
    \centering
    \includegraphics[width=1.1\textwidth]{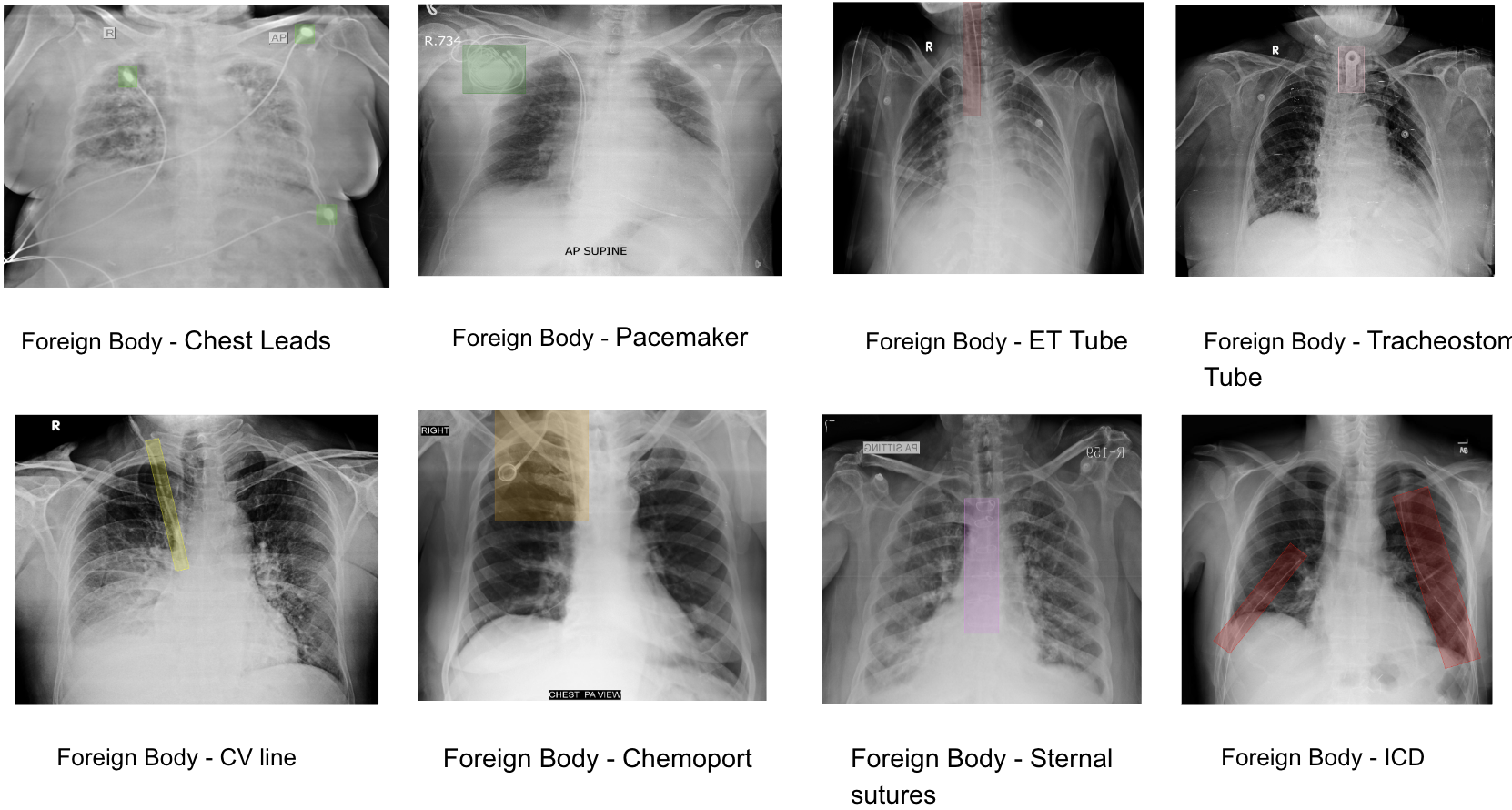} 
    \label{fig:detection4}
\end{figure}

\begin{figure}[H] 
    \centering
    \includegraphics[width=1.1\textwidth]{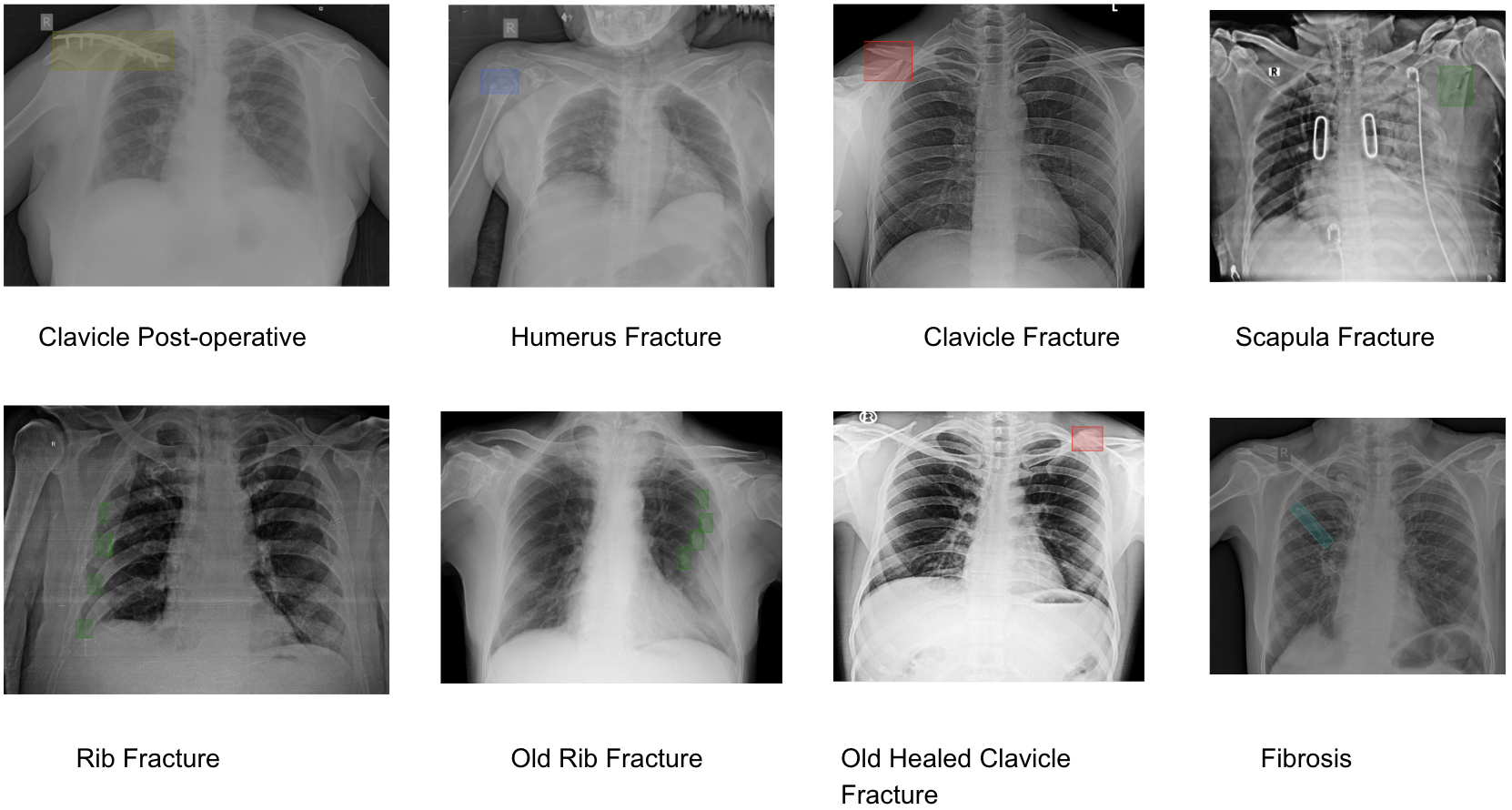} 
    \label{fig:detection5}
\end{figure}

\begin{figure}[H] 
    \centering
    \includegraphics[width=1.1\textwidth]{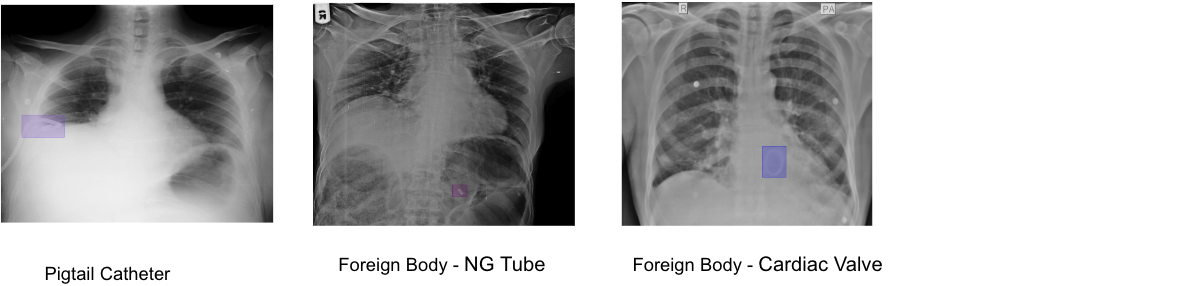} 
    \label{fig:detection6}
    \caption{Pathology Detections}
\end{figure}

\section*{Multi-Site Clinical Trial and Dataset Composition}
This study was conducted as a multi-site clinical trial across healthcare facilities in India, encompassing Government hospitals, Large Private Enterprise Hospitals(Including 17 Large Enterprise healthcare entities in India), and Small to medium-sized (SME) hospitals. The trial included a dataset of over 1 million chest X-ray scans collected from these sites, providing a robust and diverse sample for model evaluation. This dataset allowed for testing under a wide range of imaging conditions, from high-resolution scans in well-resourced private hospitals to lower-quality images often encountered in government and SME settings.
The trial aimed to assess the model's diagnostic accuracy, consistency, and reliability across diverse imaging environments. Each scan was processed through the model’s classification and detection phases, allowing us to capture detailed performance metrics such as sensitivity, specificity, precision, and recall. The multi-site setup enabled cross-validation across varying levels of image quality, patient demographics, and clinical workflows, ensuring that the model could generalize effectively. By evaluating performance across 1 million scans in different healthcare settings, this study provided a rigorous assessment of the model’s applicability and scalability for widespread deployment within the Indian healthcare system.

\subsection*{Subgroup Analysis}

Subgroup analysis is essential to assess the model’s generalizability across diverse clinical conditions and demographics. By evaluating performance across age, X-ray machine manufacturer, and gender, we ensure that the model can handle anatomical differences, equipment variability, and demographic diversity without bias. This approach verifies that the model maintains consistent accuracy, precision, and recall across real-world scenarios, making it reliable and adaptable for deployment across varied healthcare settings.

The results for accuracy, precision, recall, sensitivity, and specificity across these subgroups are presented in the table below:

\maketitle

\begin{table}[h!]
\centering
\begin{tabular}{|p{0.12\textwidth}|p{0.12\textwidth} | p{0.12\textwidth}|p{0.12\textwidth}|p{0.12\textwidth}|p{0.12\textwidth}|p{0.12\textwidth}|}
\hline
\textbf{Age Group} & \textbf{AUC} & \textbf{Accuracy (\%)} & \textbf{Precision (\%)} & \textbf{Recall (\%)} & \textbf{Sensitivity (\%)} & \textbf{Specificity (\%)} \\
\hline
Under 18 & 0.903 & 96.5 & 96.2 & 96.8 & 96.5 & 97.0 \\
\hline
18-40 & 0.986 &  98.2 & 98.0 & 98.3 & 98.0 & 98.5 \\
\hline
40-60 & 0.972 & 97.8 & 97.5 & 98.0 & 97.8 & 98.3 \\
\hline
60-75 & 0.976 & 97.2 & 96.3 & 96.8 & 97.0 & 98.1 \\
\hline
75+ & 0.885 & 95.3 & 94.8 & 95.0 & 96.3 & 96.9 \\
\hline
\end{tabular}
\caption{Performance Metrics by Age Group}
\end{table}

\begin{table}[h!]
\centering
\begin{tabular}{|p{0.12\textwidth}|p{0.12\textwidth}| p{0.12\textwidth}|p{0.12\textwidth}|p{0.12\textwidth}|p{0.12\textwidth}|p{0.12\textwidth}|}
\hline
\textbf{Gender} & \textbf{AUC} & \textbf{Accuracy (\%)} & \textbf{Precision (\%)} & \textbf{Recall (\%)} & \textbf{Sensitivity (\%)} & \textbf{Specificity (\%)} \\
\hline
Male & 0.986 & 98.0 & 97.9 & 98.1 & 98.0 & 98.2 \\
\hline
Female & 0.979 & 97.8 & 97.6 & 98.1 & 97.8 & 98.0 \\
\hline
\end{tabular}
\caption{Performance Metrics by Gender}
\end{table}

\begin{table}[h!]
\centering
\begin{tabular}{|p{0.12\textwidth}|p{0.12\textwidth}| p{0.12\textwidth}|p{0.12\textwidth}|p{0.12\textwidth}|p{0.12\textwidth}|p{0.12\textwidth}|}
\hline
\textbf{Machine Type} & \textbf{AUC} & \textbf{Accuracy (\%)} & \textbf{Precision (\%)} & \textbf{Recall (\%)} & \textbf{Sensitivity (\%)} & \textbf{Specificity (\%)} \\
\hline
CR & 0.978 & 98.2 & 97.4 & 98.3 & 97.6 & 97.9 \\
\hline
DR & 0.969 & 97.5 & 96.9 & 97.5 & 97.1 & 98.0 \\
\hline
\end{tabular}
\caption{Performance Metrics by Machine Type}
\end{table}

\begin{table}[h!]
\centering
\begin{tabular}{|p{0.18\textwidth}|p{0.12\textwidth}|p{0.12\textwidth}|p{0.12\textwidth}|p{0.12\textwidth}|p{0.12\textwidth}|p{0.12\textwidth}|}
\hline
\textbf{Manufacturer} & \textbf{AUC} & \textbf{Accuracy (\%)} & \textbf{Precision (\%)} & \textbf{Recall (\%)} & \textbf{Sensitivity (\%)} & \textbf{Specificity (\%)} \\
\hline
GE Healthcare & 0.95 & 98.1 & 97.8 & 98.3 & 98.0 & 97.5 \\
\hline
Siemens & 0.967 &  97.9 & 97.6 & 98.1 & 97.7 & 98.0 \\
\hline
Philips & 0.921 & 98.0 & 97.9 & 98.2 & 97.8 & 98.3 \\
\hline
Other Manufacturers & 0.934 &  97.5 & 97.3 & 97.8 & 97.6 & 97.7 \\
\hline
\end{tabular}
\caption{Performance Metrics by Manufacturer}
\end{table}

\section*{Deployment at live Radiology Reporting Workflow: }

The AI software has been deployed across 17 major healthcare systems in India, covering both urban and rural settings. Integrated into 5C Network's clinical workflows, the AI system has processed over 157,369 chest X-rays, averaging around 2,000+ CXR scans per day. The system performs automated classification of chest X-rays into normal and abnormal categories and detects specific pathologies, enabling rapid identification and triage of scans. This setup reduces time spent on routine cases, allowing radiologists to focus on critical findings like pneumothorax, which require prompt intervention.

\subsection*{Radiologist Validation of AI Predictions: }
After the AI processes a scan, its predictions are presented to the radiologists for review. Radiologists assess the AI-generated findings, validating or rejecting each classification and annotation. This feedback mechanism allows radiologists to indicate which predictions are accurate and identify any discrepancies, providing direct input on the model's performance. Each accepted or rejected prediction is logged, enabling continuous refinement of the AI model based on radiologists' expertise and feedback. This validation loop ensures that the AI adapts to clinical requirements and maintains high predictive accuracy.

\subsection*{Post-deployment Results: }

\begin{longtable}{|p{0.5\textwidth}|p{0.15\textwidth}|p{0.16\textwidth}|p{0.15\textwidth}|}
\hline
\textbf{Pathology} & \textbf{AUC} & \textbf{Precision (\%)} & \textbf{Recall (\%)} \\
\hline
\endfirsthead

\hline
\textbf{Pathology} & \textbf{AUC} & \textbf{Precision (\%)} & \textbf{Recall (\%)} \\
\hline
\endhead

\hline \multicolumn{4}{r}{\textit{Continued on next page}} \\ \hline
\endfoot

\hline
\caption{Performance Metrics for Detected Pathologies} \label{table:performance_metrics} \\
\endlastfoot

Alveolar Lung Opacity & 0.96 & 96.90 & 96.08 \\
\hline
Atelectasis & 0.99 & 98.11 & 97.75 \\
\hline
Azygous Lobe & 0.99 & 99.03 & 99.64 \\
\hline
Bifid Rib & 0.99 & 96.22 & 93.70 \\
\hline
Bronchiectasis & 0.98 & 99.12 & 98.88 \\
\hline
Bullous Emphysema & 0.97 & 98.52 & 94.83 \\
\hline
Cardiomegaly & 0.96 & 96.85 & 95.40 \\
\hline
Cavity & 0.97 & 98.10 & 97.40 \\
\hline
Cervical Rib & 0.96 & 96.00 & 95.43 \\
\hline
Clavicle Fracture & 0.98 & 98.52 & 95.50 \\
\hline
Clavicle Fracture with PO & 0.98 & 97.30 & 98.90 \\
\hline
Consolidation & 0.97 & 96.17 & 93.81 \\
\hline
Dextrocardia & 0.98 & 97.38 & 98.65 \\
\hline
Dextrocardia with situs inversus & 0.99 & 98.78 & 92.50 \\
\hline
Diaphragmatic Hump & 0.97 & 97.75 & 97.23 \\
\hline
Elevated Diaphragm & 0.97 & 95.42 & 99.60 \\
\hline
Esophageal Stent & 0.97 & 97.50 & 98.90 \\
\hline
Fibrosis & 0.99 & 97.61 & 99.73 \\
\hline
Fissural Thickening & 0.98 & 98.88 & 97.80 \\
\hline
Flattened Diaphragm & 0.99 & 98.20 & 100.38 \\
\hline
Foreign Body - Cardiac Valves & 1.00 & 99.71 & 99.69 \\
\hline
Foreign Body - Chemoport & 0.97 & 98.33 & 95.00 \\
\hline
Foreign Body - Chest Leads & 0.99 & 98.95 & 98.15 \\
\hline
Foreign Body - CV Line & 0.97 & 98.47 & 95.60 \\
\hline
Foreign Body - ETT & 0.98 & 99.13 & 96.99 \\
\hline
Foreign Body - ICD & 0.96 & 95.98 & 95.38 \\
\hline
Foreign Body - Nasojejunal Tube & 0.95 & 98.92 & 99.72 \\
\hline
Foreign Body - NG Tube & 0.96 & 95.80 & 95.40 \\
\hline
Foreign Body - Pacemaker & 0.99 & 98.91 & 98.93 \\
\hline
Foreign Body - Pigtail Catheter & 0.97 & 98.01 & 96.10 \\
\hline
Foreign Body - Spinal Fusion & 1.00 & 100.00 & 99.50 \\
\hline
Foreign Body - Sternal Sutures & 1.00 & 100.00 & 99.40 \\
\hline
Foreign Body - Tracheostomy Tube & 0.97 & 96.10 & 97.40 \\
\hline
Hilar Lymphadenopathy & 0.98 & 98.23 & 94.60 \\
\hline
Hilar Prominence & 0.99 & 98.81 & 98.89 \\
\hline
Humerus Fracture & 0.99 & 98.92 & 98.35 \\
\hline
Humerus Post OP & 0.98 & 98.65 & 96.88 \\
\hline
Hydro Pneumothorax & 0.99 & 98.01 & 97.83 \\
\hline
Hypoplastic Rib & 0.99 & 98.23 & 95.10 \\
\hline
ILD & 0.98 & 99.03 & 99.64 \\
\hline
Interstitial Lung Opacity & 0.98 & 98.10 & 98.18 \\
\hline
Lobe Collapse & 0.96 & 96.35 & 96.20 \\
\hline
Lung Collapse & 0.97 & 97.98 & 96.18 \\
\hline
Lung Mass & 0.97 & 97.75 & 96.85 \\
\hline
Lymph Node Calcification & 0.96 & 97.70 & 95.72 \\
\hline
Mastectomy & 0.97 & 99.03 & 98.20 \\
\hline
Mediastinal Mass & 0.96 & 96.25 & 95.28 \\
\hline
Mediastinal Shift & 0.97 & 97.60 & 96.70 \\
\hline
Mediastinal Widening & 0.98 & 98.49 & 95.57 \\
\hline
Milliary Tuberculosis & 0.98 & 98.49 & 94.72 \\
\hline
Nodule & 1.00 & 98.83 & 99.72 \\
\hline
Old Healed Clavicle Fracture & 0.98 & 98.68 & 97.88 \\
\hline
Old Rib Fracture & 0.98 & 100.38 & 99.50 \\
\hline
Old TB & 1.00 & 99.05 & 97.43 \\
\hline
Pericardial Cyst & 0.96 & 97.10 & 95.40 \\
\hline
Pleural Calcification & 0.99 & 99.70 & 96.91 \\
\hline
Pleural Effusion & 0.96 & 96.80 & 96.00 \\
\hline
Pleural Plaque & 0.98 & 98.81 & 98.11 \\
\hline
Pleural Thickening & 0.96 & 96.52 & 95.12 \\
\hline
Pneumonia & 0.97 & 99.03 & 97.52 \\
\hline
Pneumoperitoneum & 0.98 & 98.34 & 96.31 \\
\hline
Pneumothorax & 0.96 & 96.22 & 95.63 \\
\hline
Prominent Bronchovascular Markings & 0.98 & 98.10 & 98.18 \\
\hline
Pulmonary Edema & 0.95 & 96.11 & 94.38 \\
\hline
Reticulo-nodular Appearance & 0.97 & 98.92 & 97.10 \\
\hline
Rib Fracture & 0.98 & 97.92 & 98.03 \\
\hline
Scapula Fracture & 0.96 & 98.90 & 96.65 \\
\hline
Scoliosis & 0.98 & 96.80 & 96.83 \\
\hline
Subcutaneous Emphysema & 0.99 & 98.40 & 99.70 \\
\hline
Surgical Staples & 0.98 & 95.51 & 94.00 \\
\hline
Thyroid Lesion & 0.97 & 97.33 & 99.00 \\
\hline
Tracheal and Mediastinal Shift & 0.95 & 97.95 & 96.38 \\
\hline
Tracheal Shift & 0.98 & 96.51 & 98.90 \\
\hline
Tuberculosis & 0.97 & 98.88 & 100.35 \\
\hline
Unfolding of aorta & 0.98 & 98.81 & 98.90 \\
\hline

\end{longtable}

\section*{Limitations and Considerations}

Our AI model, while demonstrating substantial advancements in multi-pathology detection for chest X-rays, is subject to certain limitations and considerations:

\begin{itemize}

    \item \textbf{Complexity in Pathology Segmentation}: Accurately segmenting overlapping structures or low-contrast areas, such as dense lung regions, can be challenging, potentially affecting segmentation precision.

    \item \textbf{Dependence on Image Preprocessing}: The model’s accuracy relies on consistent preprocessing steps, and deviations in techniques, such as rotation correction or contrast adjustments, may affect detection performance.

    \item \textbf{Population-Specific Adaptability}: With training data largely sourced from Indian healthcare systems, the model’s generalizability to other populations or regions might be limited.
\end{itemize}

\section*{Conclusion}

This study demonstrates the efficacy of an autonomous AI system for multi-pathology detection in chest X-rays, addressing critical diagnostic gaps in the Indian healthcare landscape. By leveraging advanced deep-learning architectures such as Vision Transformers, Faster R-CNN, and U-Net variants, the system achieved high precision and recall across 75 pathologies, validated on over 5 million CXRs and deployed across 17 major healthcare institutions. Notably, its integration into clinical workflows has reduced radiology reporting times by up to 50\%, significantly improving efficiency while maintaining diagnostic accuracy. This reduction is particularly impactful in addressing the high demand for radiology services, enabling faster decision-making and better patient care, especially in underserved regions where radiologist shortages persist. The AI-driven approach offers a scalable, cost-effective solution that ensures consistent, high-quality interpretations, enhancing overall healthcare outcomes. Looking ahead, further advancements will focus on expanding pathology coverage, improving AI-assisted clinical decision support, and optimizing deployment in resource-limited settings. Additionally, broader multi-site validation beyond India will be essential for ensuring global applicability. As AI continues to evolve, its role in radiology will not only augment diagnostic capabilities but also redefine the future of autonomous medical imaging.

\end{document}